\documentclass[twoside, a4paper]{article}

\newcommand{\nomthm}{Theorem}
\newcommand{\nomprop}{Proposition}
\newcommand{\nomcor}{Corollary}
\newcommand{\nomdf}{Definition}
\newcommand{\nomlem}{Lemma}
\newcommand{\nomdem}{Proof}
\newcommand{\nomeg}{Example}

\usepackage[T1]{fontenc}
\usepackage[utf8]{inputenc}
\usepackage{lmodern}

\usepackage{indentfirst}

\usepackage{amsmath}
\usepackage{amssymb}
\usepackage{amsfonts}
\usepackage{stmaryrd}
\usepackage{dsfont}

\usepackage{graphicx}
\usepackage{fancybox}
\usepackage{color}
\usepackage{shadow}
\usepackage{array}
\usepackage{framed}
\usepackage{csquotes}
\usepackage{enumitem}

\usepackage{appendix}

\usepackage{pgf}
\usepackage{tikz}
\usepackage{tikz-cd}

\usepackage{algorithmic}
\usepackage{algorithm}

\usepackage{mathpartir}

\usepackage{listings}
\usepackage[textsize=small]{todonotes}

\usepackage[a4paper, 
margin=2cm
]{geometry}

\usepackage[
colorlinks=false,
urlcolor= blue,
linkcolor= red,
bookmarksopen=true]{hyperref}

\renewcommand{\emph}{\textbf}

\newcommand{\ZZ}{\mathbb{Z}}
\newcommand{\ZZn}[1]{\ZZ/n\ZZ}

\newcommand{\CC}{\mathbb{C}}
\newcommand{\DD}{\mathbb{D}}

\newcommand{\EE}{\mathbb{E}}

\newcommand{\Ee}{\mathcal{E}}
\newcommand{\Ff}{\mathcal{F}}
\newcommand{\Gg}{\mathcal{G}}
\newcommand{\Pp}{\mathcal{P}}
\newcommand{\Mm}{\mathcal{M}}

\newcommand{\Aa}{\mathcal{A}}
\newcommand{\Uu}{\mathcal{U}}
\newcommand{\Tt}{\mathcal{T}}
\newcommand{\Dd}{\mathcal{D}}

\newcommand{\zero}{\mathbf{0}}
\newcommand{\one}{\mathbf{1}}
\newcommand{\two}{\mathbf{2}}

\renewcommand{\epsilon}{\varepsilon}
\renewcommand{\phi}{\varphi}

\newcommand{\pta}[2]{\forall\:#1\in#2,\ }

\newcommand{\exa}[2]{\exists\:#1\in#2,\ }

\newcommand{\imp}{\Rightarrow}
\newcommand{\equ}{\Leftrightarrow}

\DeclareMathOperator{\id}{id}

\usepackage[style=numeric]{biblatex}
\usetikzlibrary{automata,positioning}

\usepackage[thmmarks, framed]{ntheorem}
\theorembodyfont{\rmfamily}
\theoremstyle{break}

\newframedtheorem{thm}{\nomthm}[section]
\newframedtheorem{prop}[thm]{\nomprop}
\newframedtheorem{cor}[thm]{\nomcor}
\newframedtheorem{lem}[thm]{\nomlem}

\newframedtheorem{df}{\nomdf}[section]
\newframedtheorem{eg}{\nomeg}[section]

\theoremstyle{nonumberbreak}	
\theorembodyfont{\small}

\newtheorem{dem}{\nomdem}[section]

\DeclareFontFamily{U}{mathc}{}
\DeclareFontShape{U}{mathc}{m}{it}
{<->s*[1.03] mathc10}{}
\DeclareMathAlphabet{\mathcal}{U}{mathc}{m}{it}

\DeclareMathOperator{\beh}{beh}
\DeclareMathOperator{\dom}{dom}
\DeclareMathOperator{\cod}{codom}
\DeclareMathOperator{\To}{\Rightarrow}
\DeclareMathOperator{\upar}{\uparrow}
\DeclareMathOperator{\downar}{\downarrow}
\DeclareMathOperator{\uphar}{\upharpoonleft}
\DeclareMathOperator{\downhar}{\downharpoonleft}
\DeclareMathOperator{\st}{st}

\DeclareMathOperator{\Set}{\mathbf{Set}}
\DeclareMathOperator{\Up}{\mathcal{U\mkern-3mu p}}

\DeclareMathOperator{\UUp}{\widetilde{\Up}}
\DeclareMathOperator{\Dn}{\mathcal{D\mkern-3mu n}}
\DeclareMathOperator{\Poset}{\mathbf{Poset}}
\DeclareMathOperator{\Alt}{\mathcal{A\mkern-3mu lt}}
\DeclareMathOperator{\Em}{\mathcal{EM}}

\newcommand{\aut}{\Dd}
\DeclareMathOperator{\emptyword}{\epsilon}
\DeclareMathOperator{\Do}{\mathcal{O}}

\title{Coalgebraic Determinization of Alternating Automata \\ \normalsize Report on a M1 Internship}
\author{Meven \textsc{Bertrand}, under the supervision of Jurriaan \textsc{Rot}}
\date{September 25 2017}
\hypersetup{
	pdftitle={Determinization of Alternating Automata},
	pdfauthor={Meven Bertrand},
	pdfsubject={Mathematics}
	}

\begin{document}

\maketitle

\begin{abstract}
	Coalgebra is a currently quite active field, which aims to look at generic state-based systems (most prominently automata) from a very abstract point of view, mainly using tools from category theory. One of its achievements is to give a completely generic approach of determinization, unifying in an elegant manner non-deterministic automata, probabilistic automata or non-deterministic pushdown automata in one and the same model.
	
	However, the case of alternating automata fails to easily fit in this model. The aim of this internship was therefore to tackle this problem: can alternating automata also be determinized in the coalgebraic way? Does this give semantics that coincides with the concretely defined one?
	
	In this report, we give a positive answer to both questions. The main element of our construction is a distributive law, the definition of which has been for some time an open question.
	
\end{abstract}

\tableofcontents

\section{Introduction}

A big part of computer science is about studying models of computation, for various purposes. A lot of these models (automata, Turing machines, stream systems, Mealy machines, etc.) share a common structure: they consist of a set of states, together with some kind of transition structure. The “user” does not have access to the states, but only to some output: the system behaves like a black box.

Coalgebra is a field that uses tools coming from category theory to describe systems of this kind in a generic way. Its biggest achievement is the notion of bisimulation, a sound reasoning principle for behavioural equivalence that exists for any of these state-based systems as soon as they are described in the coalgebraic setting.

However, this notion of bisimulation will not be our main concern here. Rather, we will look at another quite interesting construction: determinization. A lot of state based systems include some kind of branching structure: it can be choice between multiple transitions (non-deterministic automata, non-deterministic Turing machines, and so forth), but also probabilistic transitions (probabilistic automata), and even more exotic models (weighted automata, alternating automata, etc.). In most of these cases, one can construct a deterministic version of the system, by making the state space bigger. Thanks to the coalgebraic setting, this determinization is now understood at an abstract level, and can be performed on generic non-deterministic systems, as surveyed in \cite{TraceSem}.

However, there are a few cases that somehow do not easily fit in this setting. The case of alternating automata is one, and it has resisted attempts to describe it coalgebraically for some time. In this report we give a coalgebraic description of alternating automata in the setting of determinization. This finally solves the aforementioned difficulty, and makes alternating automata another example of the power of the coalgebraic theory. Moreover, this definition gives a semantics that is similar to the concrete one, which makes it very satisfactory.

The crucial point of this determinization is a categorical construction called distributive law. The law itself was already known, but it was used in a completely different setting. Thus, our work was mostly to acknowledge that this law could be used in our setting, and construct the frame around it to make it fit into the general picture of determinization. After this, we still had to verify that the semantics corresponding to this construction is indeed the usual, concrete one.

\paragraph{Literature review}
The search for this distributive law has caused a lot of mistakes and unsatisfactory trials: in \cite{BialgDeterm} and \cite{TraceSem}, the problem of alternating automata is tackled, but each time the model is less interesting than the original one; concerning the law itself, a construction that is erroneously claimed to be a distributive law appears in different papers, for instance \cite{ForgLog15} (corrected in \cite{ForgLog16}), or \cite{MonSet}. The case of~\cite{ForgLog16} is rather interesting, as they use an imperfect law, and are still able to draw some results out of it, but this is not enough to really solve the problem. A precise description of the errors can be found in \cite{ErrMon} (which is, to my knowledge, unpublished and unavailable online). The use of our the distributive law was suggested by Luigi Santocanale. It seems to be somewhat a folklore result, so it is hard to trace back, but it can for example be found in \cite[p.~220-221]{BasDL}, where it is just a simple example of a much more complex construction.

\paragraph{Outline}
The report is divided in three parts: the first one (section \ref{catth}) gives some generic categorical notions, the second one (section \ref{coalgdet}) introduces the needed notions of coalgebra, including the generic determinization procedure, and the last one (section \ref{casalt}) exposes the special case of alternating automata. This last section is the one containing the original work, the two others are mostly there to give context.

\section{Category Theory}
\label{catth}

In this section, we give a small overview of category theory, with the aim of introducing the main subjects the internship was dealing with. Since we mean it to be an overview, and not a proper course of any kind, we will not give proofs, but just state the interesting properties, and try to give intuition on the reason they are true. Complete definitions and proofs can be found in \cite{Awodey}.

\subsection{Basic Definitions}
\label{basobj}

In almost every field of mathematics, one is concerned with a certain type of objects, and with mappings that preserve these objects: sets and functions, vectors spaces and linear maps, groups and group homomorphisms, ordered sets and monotone maps, topological spaces and continuous functions, and so on. This is what category theory tries to define, on a very abstract level.

\begin{df}[Category]
	A category consists of
	\begin{itemize}
		\item a collection of objects ($A$, $B$, $C$ \dots),
		\item a collection of arrows ($f$, $g$, $h$ \dots),
	\end{itemize}
	and the following are also given
	\begin{itemize}
		\item for each arrow, two objects are given, its domain and codomain, written respectively as $\dom(f)$ and $\cod(f)$, and we usually write $f : A \to B$ to indicate that $\dom(f) = A$ and $\cod(f) = B$,
		\item given any three objects $A$, $B$ and $C$ and arrows $f : A \to B$ and $g : B \to C$, an arrow $g \circ f : A \to C$ is given, called the composite of $f$ and $g$,
		\item for each object, an arrow $\id_A$ is given, called the identity of $A$,
	\end{itemize}
	and these must satisfy
	\begin{itemize}
		\item associativity: $h \circ (g \circ f) = (h \circ g) \circ f$ whenever the composition is defined,
		\item unit: for any arrow $f : A \to B$, we have $\id_B \circ f = f = f \circ \id_A$.
	\end{itemize}
\end{df}

All the examples mentioned above are categories according to this definition. Some much weirder constructions are also categories, but we will not present these, as they are of little interest for this report.

\paragraph{}
We will mostly work in the following category:

\begin{df}[Category of sets]
	The sets (as objects) and functions (as arrows), with the usual compositions and identities, form a category, that we will denote by $\Set$.
\end{df}

Another category we will make use of is the following:

\begin{df}[Category of posets]
	A poset (partially ordered set) is a set together with an order relation (transitive, reflexive and antisymmetric relation).\\
	A monotone map $f$ between two posets $(X,\leq)$ and $(Y,\preceq)$ is a function $f : X \to Y$ such that 
	\[\pta{x,x'}{X} x \leq x' \imp f(x) \preceq f(x')\]
	The posets and monotone maps (with the usual composition and identity) form a category, that we will denote as $\Poset$.
\end{df}

Of course, we want maps between categories. This is possible using the next definition.
\begin{df}[Functor]
	Given two categories $\CC$ and $\DD$, a functor $\Ff : \CC \to \DD$ is a mapping from objects of $\CC$ to objects of $\DD$ and from arrows of $\CC$ to arrows of $\DD$ that preserves domains, codomains, composition and identities, that is:
	\begin{itemize}
		\item for any arrow $f : A \to B$ in $\CC$, $\Ff(f) : \Ff(A) \to \Ff(B)$,
		\item for any two arrows $f : A \to B$ and $g : B \to C$, $\Ff(g \circ f) = \Ff(g) \circ \Ff(f)$,
		\item for any object $A$, $\Ff(\id_A) = \id_{\Ff(A)}$.
	\end{itemize}
\end{df}

This definition can apply to an extremely large amount of contexts and usual constructions, and this is the main reason why category theory is so handy in formulating coalgebra, as we will see in section~\ref{sbsascoalg}.

\paragraph{}
Here are some examples that we will use later in this report.

\begin{eg}[Product as a functor]
	Given a set $X$, one can define the Cartesian product with $X$ as a functor $\Ff$ in the category $\Set$:
	\begin{itemize}
		\item on objects, we define $\Ff(Y) = X \times Y$,
		\item on arrows, we define $\Ff(f)(x,y) = (x, f(y))$, that is, $\Ff(f)$ is the identity on the $X$ component, and $f$ on the other component.
	\end{itemize}
	One can easily check that this fits into the definition above. This functor will be denoted as $X \times (-)$ later in this report.
\end{eg}

\begin{eg}[Exponential as a functor]
	Given a set $X$, one can define the exponential with respect to $X$ as a functor $\Ff$ in the category $\Set$:
	\begin{itemize}
		\item on objects, we define $\Ff(Y) = Y^X$, the set of functions from $X$ to $Y$,
		\item on arrows, we define $\Ff(f)(g) = f \circ g$, that is $\Ff(f)$ is the post-composition with $f$.
	\end{itemize}
	Again, it is easy to check that this fits into the definition above. This functor will be denoted as $(-)^X$ later in this report.
\end{eg}

\begin{eg}[Identity functor]
	Given a category $\CC$, there is a so-called identity functor on $\CC$, written as $\id_\CC$, defined as the identity both on objects and arrows.
\end{eg}

\begin{eg}[Powerset functor]
	The usual powerset construction can be seen as a functor $\Pp : \Set \to \Poset$, defined as follows:
	\begin{itemize}
		\item on objects, we define $\Pp(X)$ as the set of all subsets of $X$, with the order given by inclusion,
		\item on arrows, we define $\Pp(f)(P) = \{f(x), x \in P \}$, that is $\Ff(f)$ takes the direct image of a subset by $f$.
	\end{itemize}
	If one forgets about the order structure, one can also see the same $\Pp$ as a functor $\Pp : \Set \to \Set$. In the rest of the report, we will abuse notation and make no distinction between both functors, as the correct type can be inferred from the context.
\end{eg}

\begin{eg}[Forgetful functor]
	A large and very useful class of functors is the class of forgetful functors, that is functors between categories that “forget” some of the structure. For instance, we can look at the category $\Poset$. Its objects are ordered sets, but one can “forget” that these objects have some order structure, and rather look at them purely as sets. That is what the following forgetful functor $\Uu$ does:
	\begin{itemize}
		\item on objects, we define $\Uu((X,\leq)) = X$,
		\item on arrows, we define $\Uu(f) = f$ (but $f$ which was a monotone function is now considered as a mere function between sets).
	\end{itemize}
\end{eg}

A last definition, which characterize to objects that are in a certain sense “universal”.
\begin{df}[Final object]
	A final object of a category $\CC$ is an object $\one$ of $\CC$ such that for any other object $A$ of $\CC$, there exists a unique arrow $!_A : A \to \one$.
\end{df}

\subsection{More Advanced Constructions}
\label{advconstr}

Until here, the constructions we defined were rather simple, and though they may be useful as such, the real interest in category theory comes from more complex constructions. These constructions are also the ones that are at the core of the coalgebraic setting we will work in, mostly the notion of monad, that we will use to encapsulate non-determinism.

The definitions will be given fully and illustrated with examples, in order to give some grasp on them, but the details of the definitions are not needed to understand the rest of the report past this section.

\begin{df}[Natural transformation]
	Given two functors $\Ff : \CC \to \DD$ and $\Gg : \CC \to \DD$, a natural transformation $\lambda : \Ff \To \Gg$ is a family of arrows (in $\DD$) $\lambda_X : \Ff(X) \to \Gg(X)$ (one for each object $X$ of $\CC$) such that for any arrow $f : X \to Y$ in $\CC$, the following diagram commutes:\\
	\begin{tikzcd}
		\Ff(X) \arrow[d, "\Ff(f)"'] \arrow[r, "\lambda_X"] & \Gg(X) \arrow[d, "\Gg(f)"] \\
		\Ff(Y) \arrow[r, "\lambda_Y"'] & \Gg(Y)
	\end{tikzcd}\\
	that is, such that $\lambda_Y \circ \Ff(f) = \Gg(f) \circ \lambda_X$.
\end{df}

The idea of a natural transformation is that it is a transformation between two functors that can be indifferently applied before of after a function application. In a sense, it transforms $\Ff(X)$ into $\Gg(X)$ without looking at the content of $X$, but only at the structure given by $\Ff$ and $\Gg$.

\begin{eg}[Cartesian Product]
	Recall the functor $X \times (-)$ that we already defined on $\Set$. We could also define in a very similar way the functor $(-) \times X$. Then there is a natural transformation $\lambda : X \times (-) \To (-) \times X$ between these two functors, defined by the components:
	\[\begin{array}{rccl}
	\lambda_Y :& X \times Y & \to & Y \times X \\
	& (x,y) & \mapsto & (y,x)
	\end{array}\]
\end{eg}

This notion of natural transformation is already useful as such, but it is also a building element in a lot of other categorical notions, such as the one of monad.

\begin{df}[Monad]
	A monad is a functor $\Tt$ from a category $\CC$ to itself, together with two natural transformations $\eta : \id \To \Tt$ and $\mu : \Tt \circ \Tt \to \Tt$ such that the following diagrams commute for any object $X$:\\
	\begin{tikzcd}
		\Tt(X) \arrow[r,"\eta_{\Tt(X)}"] \arrow[dr,"\id_X"] \arrow[d,"\Tt(\eta_X)"'] & \Tt(\Tt(X)) \arrow[d,"\mu_X"] \\
		\Tt(\Tt(X)) \arrow[r,"\mu_X"'] & \Tt(X)
	\end{tikzcd}
	\begin{tikzcd}
		\Tt(\Tt(\Tt(X))) \arrow[d,"\mu_{\Tt(X)}"'] \arrow[r,"\Tt(\mu_{X})"] & \Tt(\Tt(X)) \arrow[d,"\mu_X"] \\
		\Tt(\Tt(X)) \arrow[r,"\mu_{X}"'] & \Tt(X)
	\end{tikzcd}\\
	that is we have $\mu_X \circ \eta_{\Tt(X)} = id_X = \mu_X \circ \Tt(\eta_X)$ and $\mu_X \circ F(\mu_X) = \mu_X \circ \mu_{F(X)}$.\\
	The transformation $\eta$ is called the unit of the monad, and the transformation $\mu$ is called its multiplication.
\end{df}

The idea is that a monad is some kind of structure, along with a way to put an object into this structure (the unit), and a way to collapse two levels of the structure together. These must respect some easy rules: if you use the unit to put a second level, and collapse this level, then you get the thing you started with, and if you collapse three levels into one, you can collapse the outer most or inner most two first, with the same result.

\begin{eg}[Powerset as a monad]
	We already described the powerset as a functor $\Pp$ from the category $\Set$ to itself. This functor can be made into a monad, by taking:
	\begin{itemize}
		\item the unit $\begin{array}{rccl}
		\eta_X :& X & \to & \Pp(X) \\
		& x & \mapsto & \{ x\}
		\end{array}$,
		\item the multiplication $\begin{array}{rccl}
		\mu_X :& \Pp(\Pp(X))& \to & \Pp(X) \\
		& S & \mapsto & \cup S = \cup_{s \in S} s
		\end{array}$ (that is, take a set of sets to their union).
	\end{itemize}
\end{eg}

The last important definition is the one of adjoints. This is a quite difficult notion to grasp, although it is very powerful by itself. However, in this report we will mostly be interested in their link with monads, as adjunctions are a usual way to construct monads.

\begin{df}[Adjoints]
	Two functors $\Ff : \CC \to \DD$ and $\Gg : \DD \to \CC$ are said to be adjoints (this is written as $\Ff \dashv \Gg$, $\Ff$ is called the left adjoint and $\Gg$ is called the right adjoint) if there are two natural transformation $\eta : \id_\CC \To \Gg \circ \Ff$ and $\epsilon : \Ff \circ \Gg \To \id_\DD$ such that for any objects $C$ of $\CC$ and $D$ of $\DD$, we have
	\[\Gg(\epsilon_{D}) \circ \eta_{\Gg(D)} = \id_{\Gg(D)} \]
	and
	\[\epsilon_{\Ff(C)} \circ \Ff(\eta_{C}) = \id_{\Ff(C)} \]
	that is, such that the two following diagrams commute:\\
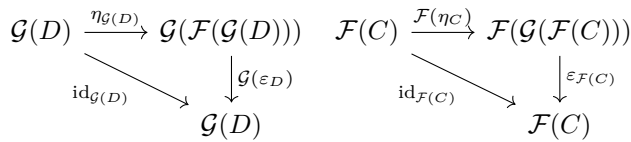
	\begin{tikzcd}
		\Gg(D) \rar["\eta_{\Gg(D)}"] \drar["\id_{\Gg(D)}"'] & \Gg(\Ff(\Gg(D))) \dar["\Gg(\epsilon_{D})"] \\
		& \Gg(D)
	\end{tikzcd}
	\begin{tikzcd}
		\Ff(C) \rar["\Ff(\eta_C)"] \drar["\id_{\Ff(C)}"'] & \Ff(\Gg(\Ff(C))) \dar["\epsilon_{\Ff(C)}"] \\
		& \Ff(C)
	\end{tikzcd}
\end{df}

There are many equivalent ways to define adjunctions. This one is good for our purposes, as it emphasizes the role of the unit and counit, which we will use in section~\ref{altaut} to define a monad. Although this definition might not seem of importance, adjoints, together with monads, form two of the most interesting objects of category theory: a lot of mathematical interesting propositions and definitions can be seen as instances of adjunctions, be it formal polynomials, quantifiers, interior and closure operators in topology, free objects over a set…

\paragraph{}
Here is an example of adjunction, of which we will make use later to construct a monad.
\begin{df}[Discrete order functor]
	The functor $\Do$ is the functor from the category $\Set$ to the category $\Poset$ that maps
	\begin{itemize}
		\item a set $X$ to the discrete order on $X$, that is $\Do(X) = (X, =)$ (in the discrete order, the only comparable elements are the ones that are equal),
		\item an arrow $f$ to itself.
	\end{itemize}
\end{df}

Note that if $f : X \to Y$ is a function in $\Set$, then $f$ is also a monotone function between $\Do(X)$ and $\Do(Y)$, so that this definition makes sense.

\begin{prop}[Adjunction for $\Do$]
	There is an adjunction $\Do \dashv \Uu$ between the discrete order functor we just defined and the forgetful functor $\Uu : \Poset \to \Set$ as defined in section~\ref{basobj}.
	
	Its unit is just the identity, and its counit is defined on a preorder $X = (S,\leq)$ by
	\[\begin{array}{rccl}
		\epsilon_X : & \Do(\Uu(X)) & \to & X \\
		& x & \mapsto & x
	\end{array}\]
	so it is the identity of $S$, but it goes from the preorder $(S,=)$ to the preorder $(S,\leq)$.
\end{prop}

\section{Coalgebras And Determinization}
\label{coalgdet}

In this section, we present the main subject of the internship, namely coalgebra and the view it gives on determinization. Once again, we do not provide the proofs of the facts we state, but these can be found in the papers we refer to.

\subsection{State-Based Systems as Coalgebras}
\label{sbsascoalg}

In this subsection we define our core construction: coalgebras. We also show how final objects can be used to give some sort of semantics. A more complete approach (with proofs!) can be found in \cite{IntCoalg}.

\begin{df}[Coalgebra]
	Given a category $\CC$, a coalgebra for a functor $\Ff : \CC \to \CC$ consists of an object $X$ together with an arrow $f : X \to \Ff X$.
\end{df}

Coalgebras are meant to represent state-based system, where $X$ represents the states of the system, and $f$ is the transition function. Then the functor $\Ff$ characterizes the type of system we are looking at.

\begin{eg}[Stream system]	
	A stream system over an alphabet (set of letters) $A$ is a set of states $X$ together with two functions $o : X \to A$ (output) and $t : X \to X$ (transition). It is a coalgebra for the functor $A \times (-)$, the transition function being $\langle o, t \rangle : X \to A \times X$, where $\langle o, t \rangle$ is defined by:
		$\begin{array}{rccl}
		\langle o, t \rangle :& X & \to & A \times X \\
		& x & \mapsto & (o(x),t(x))
		\end{array}$
\end{eg}

\begin{eg}[Deterministic automaton]
	A deterministic automaton over an alphabet $A$ is a set of states $Q$, together with a function $o : X \to \two$ (where $\two$ is the two-element set $\{\zero, \one\}$, returning $\one$ if the state is accepting and $\zero$ otherwise) and a transition function $\delta : Q \to Q^A$. It is a coalgebra for the functor $\two \times (-)^A$, the transition function being $\langle o, \delta \rangle$ where $\chi_F$.
	
	This way of looking at things is a bit unusual, but this is really the same as the usual definition of non-deterministic automata: the function $o$ corresponds to the characteristic function of the set of final states, and the function $\delta$ can be seen as a function of the type $X \times A \to X$, as for any $(x,a) \in X \times A$, $\delta(x)(a)$ is an element of $X$ (the $\delta$ of the coalgebraic definition is the curried version of the usual one).
	
	Since we will reuse this $\two \times (-)^A$ functor, we will denote it as $\aut$.
\end{eg}

Note that in the above examples we are not taking into account any starting state. This is because the aim is really to describe a state-based \emph{system}, not just one “computation” into it (for instance, the computation of a stream for a stream system, or of the acceptance or rejection of a word in an automaton).

\begin{df}[Homomorphism of coalgebras]
	Given two coalgebras $(X,f)$ and $(Y,g)$ for a functor $\Ff$, a coalgebra homomorphism is an arrow $h : X \to Y$ such that the following diagram commutes:
	\begin{tikzcd}
		X \arrow[d, "f"'] \arrow[r, "h"] & Y \arrow[d, "g"] \\
		\Ff X \arrow[r, "\Ff h"'] & \Ff Y
	\end{tikzcd}
\end{df}
That is, we can either make a transition, and then use the arrow, or use the arrow first, and then make a transition.

\paragraph{}
With this we can construct a new category.
\begin{prop}
	Given a (fixed) functor, the coalgebras and homomorphism of coalgebras for this functor form a category.
\end{prop}

\begin{df}[Final coalgebra]
	Given a functor $\Ff$, a coalgebra $(Z,\xi)$ is called a final coalgebra if it is a final object in the category of coalgebras. This means that given a coalgebra $(X,f)$ for $\Ff$, there is a unique arrow $\beh_f$ (we might call it simply $\beh$ if there is no possible confusion) such that the following diagram commutes:\\
		\begin{tikzcd}
			X \arrow[d, "f"'] \arrow[r, "\beh"] & Z \arrow[d, "\xi"] \\
			\Ff X \arrow[r, "\Ff \beh"'] & \Ff Z
		\end{tikzcd}
\end{df}

	This definition is crucial, as it gives semantics for states: if we work in the category of sets, and $x$ is an element of $X$, then $\beh_f(x)$ is the semantics of the state $x$, as the following examples illustrate.

\begin{eg}[Final stream system]
	Given an alphabet $A$, the coalgebra $(A^\omega, \langle o_z, t_z \rangle)$, where $o_z(s) = s(0)$ (the output is the first element of the stream) and $t_z(s)(i) = s(i + 1)$ (the transition is a shift of the stream), is a final coalgebra for the functor $A \times (-)$ of stream systems. Moreover, if $(X,\langle o,t \rangle)$ is a stream system, then $\beh(x) = (o(x), o(t(x)), o(t^2(x)), o(t^3(x)), \dots)$.
	
	This corresponds to the stream one would naturally associate with $x$ in the stream system $(X,\langle o, t \rangle)$.
\end{eg}

\begin{eg}[Final deterministic automaton]
	Given an alphabet $A$, the coalgebra $(\two^{A^*}, \langle O, D \rangle)$ where $\two^{A^*}$ is the set of languages over $A$,
	\[O(L) = \left\{\begin{array}{l}
	\zero \text{ if } \emptyword \notin L \\
	\one \text{ if } \emptyword \in L
	\end{array} \right.\]
	with $\emptyword$ denoting the empty word, and
	\[D(L)(a) = \{ w \in A^* \mid a \cdot w \in L \} \]
	is a final coalgebra for $\two \times (-)^A$.
	
	As expected, if $(Q, \langle o, \delta \rangle)$ is an automaton and $q$ is a state of $Q$, then $\beh(q)$ is the language accepted by the automaton with starting state $q$, as defined usually:
	\begin{itemize}
		\item $\emptyword \in \beh(q)$ if and only if $o(q) = 1$,
		\item for any word $w$ and letter $a$, $a \cdot w \in \beh(q)$ if and only if $w \in \beh(\delta(q)(a))$.
	\end{itemize}
\end{eg}

\subsection{The Problem Of Determinization}
\label{probdet}

The main subject of this internship is the determinization procedure for an automaton.
There is a well-known example, namely the one of non-deterministic automata. In that case, a new (deterministic) automaton is created, whose states are sets of states of the original automaton, and the semantics (in this case, the language denoted by the automaton) is preserved by the construction.

\begin{df}[Non-deterministic automaton]
	Given an alphabet $A$, a non-deterministic automaton consists of a set of states $Q$ together with a function $o : Q \to \two$ and a transition function $\delta : Q \to \Pp(Q)^A$. It is a coalgebra for the functor $\two \times (\Pp(-))^A$.\\
	A word $w$ is accepted by a state $q \in Q$ if
	\begin{itemize}
		\item $w = \emptyword$ and $o(q) = 1$,
		\item $w = a \cdot w'$ and $w'$ is accepted by one of the states in $\delta(q)(a)$.
	\end{itemize}
\end{df}

\begin{df}[Determinization of an automaton]
	Given a non-deterministic automaton $(Q,o,\delta)$ over an alphabet $A$, its determinization is a deterministic automaton, with
	\begin{itemize}
		\item state set $\Pp(Q)$,
		\item accepting function $o^{\#}$ defined by $o^{\#}(S) = 1 \equ \exa{s}{S} o(s) = 1$, a subset of $Q$ is accepting if it contains an accepting state,
		\item transition function $\delta^{\#} : \Pp(Q) \to \Pp(Q)^A$, defined by $\delta^{\#}(S)(a) = \cup_{q \in P} \delta(q)(a)$.
	\end{itemize}
\end{df}

\begin{thm}[Preservation of the semantics]
	\label{pressem}
	A state $q$ of a non-deterministic automaton accepts a word if and only if the state $\{q\}$ of the determinization of the automaton accepts the word.
\end{thm}

Following this motivating example, we wand to find a general way to perform the same kind of determinization. But first, we need to define what we wish to determinize.

\begin{df}[$\Tt$-automaton]
	Given a monad $(\Tt, \eta, \mu)$ on the category of sets, a $\Tt$-automaton for an alphabet $A$ is a coalgebra for the functor $\two \times (\Tt(-))^A$.
\end{df}

Here the monad represents the “non-determinisic” component of the automaton, as the example of non-deterministic automata illustrates. The $\two \times (-)^A$, as we already saw, corresponds to the automaton structure.

\begin{eg}[Non-deterministic automaton as a $\Pp$-automaton]
	A non-deterministic automaton is a $\Pp$-automaton, that is, a coalgebra for the functor $\two \times (\Pp(-))^A$.
\end{eg}

But for instance the case of probabilistic automata also fits into this definition, with a proper monad (whose functorial component involves probability distributions).

\paragraph{}
The aim is then, given a monad $\Tt$ and a $\Tt$-automaton $(X,f)$, to find a semantics for $(X,f)$, that is a function from $X$ to the final coalgebra $\two^{A^*}$, that should arise “naturally” in some sense, and correspond with the concrete semantics that one already has on the motivating examples like the non-deterministic automata.

The next two sections provide two different ways to do this. The first one gives some semantics directly, without really resorting to determinization as such. The second one is the direct abstraction of the determinization of non-deterministic automata, in the sense that from a $\Tt$-automaton with state space $X$ it constructs a deterministic automaton with state space $\Tt(X)$.

\subsection{Bialgebraic Semantics}
\label{bialg}

The construction presented here can be found in all details in \cite{BialgDeterm}. Its main ingredient is a so called Eilenberg-Moore algebra, that is an arrow $\beta : \Tt(\two) \to \two$ satisfying some axioms.

\begin{df}[Eilenberg-Moore algebra]
	Given a monad $(\Tt, \eta, \mu)$, an Eilenberg-Moore algebra is an object $X$ together with an arrow $\beta : \Tt(X) \to X$ such that the following diagrams commute:\\
	\begin{tikzcd}
		X \arrow[r, "\eta_X"] \arrow[rd, "\id_X"'] & \Tt(X) \arrow[d,"\beta"] \\
		& X
	\end{tikzcd}
	\begin{tikzcd}
		\Tt(\Tt(X))) \arrow[d, "\Tt(\beta)"'] \arrow[r, "\mu_{X}"] & \Tt(X) \arrow[d,"\beta"] \\
		\Tt(X) \arrow[r, "\beta"'] & X
	\end{tikzcd}
\end{df}

Given such an Eilenberg-Moore algebra, one can construct a new algebra on functions, using the so-called strength of the monad.

\begin{df}[Strength]
	Given a monad $\Tt$, the strength $\st : \Tt(X^Y) \to \Tt(X)^Y$ is defined by  $\st(f)(y) = \Tt(\lambda h \in X^Y \cdot h(y))(f)$.
\end{df}

Note that this construction only works because in the category $\Set$, the function $\lambda h \in X^Y \cdot h(y)$ always exists, no matter what $y$ is chosen, which is not the case in all categories.

We will use this strength operator (which actually is a natural transformation $\st : \Tt(X^Y) \To \Tt(X)^Y$) again later, as it is a very useful construction: it allows to put a monad “inside” a set of functions.

\paragraph{}
Now we can construct our algebra on functions.
\begin{prop}[Pointwise Eilenberg-Moore algebra]
	Given a Eilenberg-Moore algebra $(X,\beta)$ for a monad $\Tt$ in the category of sets and a set $Y$, one can define a new Eilenberg-Moore algebra $(X^Y, \hat{\beta})$, defined, given $f \in \Tt(X^Y)$, by
	\[\begin{array}{rccl}
	\hat{\beta}(f): & Y & \to & X \\
	& y & \mapsto & \beta(\st(f)(y))
	\end{array}\]
\end{prop}

\paragraph{}
With this algebra, we can state the main theorem of this section, found in \cite{BialgDeterm}.
\begin{thm}[Bialgebraic semantics]
	Given some monad $(\Tt, \eta, \mu)$, an Eilenberg-Moore algebra $ \beta : \Tt(\two) \to \two$, and a $\Tt$-automaton $(X,f)$, there exists a unique map $\beh_1 : X \to \two^{A^*}$ that makes the following diagram commute:\\
	\begin{tikzcd}[column sep = huge]
		X \arrow[r, "\beh_1"] \arrow[dd, "f"] & \two^{A^*} \arrow[d, "\langle O {,} D \rangle"] \\
		& \Dd(\two^{A^*})  \\
		\two \times (\Tt(X))^A = \Dd(\Tt(X)) \arrow[r, "\Dd(\Tt(\beh_1))"] & \Dd(\Tt(\two^{A^*})) \arrow[u,"\Dd(\hat{\beta})"]
	\end{tikzcd}\\
	(recall that $\Dd$ is the functor for deterministic automata, that is $\Dd = \two \times (-)^A$).
\end{thm}

This means that given only an Eilenberg-Moore algebra on $\two$, we can get a semantics for any $\Tt$-automaton, and moreover this semantics is the only one “compatible” with the algebra. When one considers in detail the above diagram for $\beh_1$, this “compatibility” says that one can either make a transition in the automaton, then go to the semantics world and use $\beta$ to aggregate the states together, or go to the semantics world and make the transition there, and that these two yield the same result.

When looking at examples, spelling out this diagram in concrete terms usually corresponds to an inductive definition of acceptance of a word, similar to the one we gave for non-deterministic automata in section \ref{probdet}: there is a rule for acceptance of the empty word, and a rule for acceptance of a word $a \cdot w$, based on the transitions made from the current state using $a$. We will have another example of this with alternating automata once we have a suitable $\Tt$-automaton structure for it.

\paragraph{}
In this theorem, the algebra $\beta$ should be seen as the formal translation of the accept condition after one step. The following example, that uses the same monad but different algebras, illustrates that.

\begin{eg}[Universal and existential non-deterministic automata]
	Given a $\Pp$-automaton, one has two “natural” ways to define $\beta : \Pp(\two) \to \two$: one can take $\beta(P) = \max(P)$ or $\beta(P) = \min(P)$.
	
	The map $\beh_{\max}$ obtained by the first one is the usual semantics, where we require that \emph{there exists} a transition $q \stackrel{a}{\to} q'$ where $q'$ accepts $w$ for $q$ to accept $a \cdot w$. This is the usual definition of a non-deterministic automaton (as appearing in section~\ref{probdet}), but we could also call it an existential non-deterministic automaton.
	
	On the contrary, the map $\beh_{\min}$ obtained by the second one corresponds to the semantics where we require that \emph{all} transitions $q \stackrel{a}{\to} q'$ lead to a state $q'$ accepting $w$ for $q$ to accept $a \cdot w$. This is also called a universal non-deterministic automaton.
\end{eg}

\subsection{Semantics Via Determinization}
\label{semdet}

The content of this section is drawn out of \cite{TraceSem}. Its main ingredient is a so called distributive law.

\begin{df}[Distributive law]
	Given a monad $(\Tt,\eta, \mu)$ and a functor $\Gg$, a distributive law is a natural transformation $\lambda : \Tt \Gg \To \Gg \Tt$, that is compatible with the monad structure, that is such that the following diagrams commute for any object $X$:\\
	\begin{tikzcd}
		\Gg(X) \arrow[r, "\eta_{\Gg(X)}"] \arrow[dr, "\Gg(\eta_X)"'] & \Tt(\Gg(X)) \arrow[d, "\lambda_X"] \\
		& \Gg(\Tt(X))
	\end{tikzcd}
	\begin{tikzcd}
		\Tt^2(\Gg(X)) \arrow[r, "\Tt(\lambda_X)"] \arrow[d, "\mu_{\Gg(X)}"'] & \Tt(\Gg(\Tt(X))) \arrow[r, "\lambda_{\Tt(X)}"] & \Gg(\Tt^2(X)) \arrow[d, "\Gg(\mu_X)"] \\
		\Tt(\Gg(X)) \arrow[rr, "\lambda_X"'] && \Gg(\Tt(X))
	\end{tikzcd}
\end{df}

\begin{thm}[Determinization via a distributive law]
	Given a monad $(\Tt, \eta, \mu)$, a functor $\Gg$, a distributive law $\lambda$ and a $\Gg \Tt$-coalgebra $(X,f)$, one can construct a determinization of $(X,f)$, namely the $\Gg$-coalgebra $(\Tt(X), \Ff_{\Ee\Mm}(f))$, where
	\[\Ff_{\Ee\Mm}(f) : \Tt(X) \to \Gg(\Tt(X)) = \left(\Tt(X) \stackrel{\Tt(f)}{\to} \Tt(\Gg(\Tt(X))) \stackrel{\lambda_{\Tt(X)}}{\to} \Gg(\Tt^2(X)) \stackrel{\Gg(\mu_X)}{\to} \Gg(\Tt(X)) \right) \]
	Moreover, this determinized coalgebra makes the following diagram commute:\\
	\begin{tikzcd}
		X \arrow[r, "\eta_X"] \arrow[d, "f"'] & \Tt(X) \arrow[dl, "\Ff_{\Ee\Mm}(f)"] \\
		\Gg(\Tt(X))
	\end{tikzcd}
\end{thm}

This commutative triangle is also called the generalized powerset construction, and has appeared for the first time is \cite{GenerDet}.

\begin{cor}
	In the context of the theorem above, if the functor $\Gg$ has a final coalgebra $(Z,\xi)$, one gets a semantics for $f$ via the unique arrow $\beh_2 : \Tt(X) \to Z$, that is the unique arrow making the following diagram commute:\\
	\begin{tikzcd}[row sep = large]
		X \arrow[r, "\eta_X"] \arrow[d, "f"] & \Tt(X) \arrow[dl, "\Ff_{\Ee\Mm}(f)"] \arrow[r, "\beh_2"] & Z \arrow[d, "\xi"]\\
		\Gg(\Tt(X)) \arrow[rr, "\Gg(\beh_2)"] && \Gg(Z)
	\end{tikzcd}
\end{cor}

In the case where $\Gg$ is the functor for automata, the situation is even better, as the distributive law can be constructed from a $\Tt$-algebra, similar to the one used in section~\ref{bialg}, as shown in \cite{TraceSem}.

\begin{prop}[Distributive law arising from an algebra]
	Given a monad $(\Tt, \eta, \mu)$ and a $\Tt$-algebra $\beta : \Tt(\two) \to \two$, there is a distributive law $\lambda$ between $\Tt$ and $\aut$, constructed as follows:
	\[\lambda_X : \Tt(\two \times X^A) \to \two \times \Tt(X)^A = \left(\Tt(\two \times X^A) \stackrel{\langle \Tt(\pi_1), \Tt(\pi_2) \rangle}{\longrightarrow} \Tt(\two) \times \Tt(X^A) \stackrel{\beta \times st}{\longrightarrow} \two \times \Tt(X)^A \right) \]
	where $\pi_1$ and $\pi_2$ are the two projection from the product $\two \times X^A$.
\end{prop}

So as before, given only a $\Tt$-algebra on $\two$ we are able to fully define a semantics.

\begin{eg}[Correspondance with concrete determinization]
	In the case where the monad is $\Pp$, the functor is $\aut$ and $\beta$ is $\max$, the obtained determinized coalgebra is exactly the same as in the usual determinization: if $(Q, \langle o, \delta \rangle )$ is a non-deterministic automaton, then $\Ff_{\Ee\Mm}(\langle o, \delta \rangle ) = \langle o^{\#}, \delta^{\#} \rangle$, with $o^{\#}$ and $\delta^{\#}$ defined as in~\ref{probdet}.
	
	Then the theorem of preservation of semantics (theorem~\ref{pressem}) just states that the semantics defined by $\beh_2 \circ \eta_Q$ is the same as the one defined concretely, because for $\Pp$, the unit $\eta$ is defined by $\eta_Q(q) = \{q\}$.
\end{eg}

The most interesting fact, however, is the following, which appears in \cite{BialgDeterm}, but relies on more high-level results from \cite{Bartels}.
\begin{thm}[Correspondance of the two semantics]
	Given a monad $(\Tt,\eta,\mu)$ and a $\Tt$-algebra $\beta : \Tt(\two) \to \two$, the two semantics $\beh_1$ of section~\ref{bialg} and $\beh_2$ are such that $\beh_2 \circ \eta_X = \beh_1$.
\end{thm}

In other words, the bialgebraic semantics and the semantics via determinization are essentially the same, even if the way they are presented is quite different. This can be interpreted in two different ways: one could argue the fact that those two constructs yield in the end the same result, is an argument in favor of thinking that they represent the “natural” way to associate a semantics to a $\Tt$-automaton.

The second way, and maybe more interesting way, is to view the previous theorem as stating that the determinization procedure yields the only semantics that is compatible with $\beta$ (in the sense we developed at the end of section~\ref{bialg}), so that it is in a way the only correct way to determinize with respect to $\beta$.

The two different constructions also have different interests: $\beh_1$ is useful in getting a semantics, and obtaining in concrete terms the definition of this semantics, while $\beh_2$ is a way to construct a new automaton, that recognizes the same language as the $\Tt$-automaton we were considering.

\section{The Case Of Alternating Automata}
\label{casalt}

\subsection{Alternating Automata}
\label{altaut}

In this section, we present the state-based system we are interested in, the alternating automaton. This model is a kind of extension of the non-deterministic automaton: in a non-deterministic automaton, a word $a\cdot w$ is accepted from a state $q$ if there is a transition $q \stackrel{a}{\to} q'$ to a state $q'$ that accepts the word $w$. However, one could choose a different rule, for instance say that \emph{every} transition labeled with $a$ should lead to a state accepting $w$. These are the simplest examples, but one could wish to use other logical rule to relate acceptation by a state to acceptation by other states. This is the idea that lead to the model of alternating automata.

Alternating automata were first introduced in \cite{Altern} with slightly more general features than the ones we consider, and in a quite different presentation. Our model has been chosen over the original one because of the ease to translate it in a categorical fashion.

\begin{df}[Alternating automata (preliminary)]
	An alternating automaton with respect to an alphabet $A$ is a coalgebra for the functor $\aut \circ \Pp \circ \Pp$, that is a set $Q$ together with a function $\langle o , \delta \rangle : Q \to \two \times \Pp(\Pp(Q))^A$. As for a deterministic automaton, the function $o : Q \to \two$ represents the accepting states. Concerning the transition function, $\Pp(\Pp(Q))$ is seen as a set of “forks”.\\
	Acceptation of a word is then defined by induction, as follows:
	\begin{itemize}
		\item a state $q$ accepts the empty word if and only if $o(q) = \one$,
		\item a state $q$ accepts the word $a \cdot w$ if and only if there is a fork in $\delta(q)(a)$ such that every state of the fork accepts $w$, that is $accepts(q,a \cdot w) \equ \exa{F}{\delta(q)(a)} \pta{q'}{F} accepts(q,w)$.
	\end{itemize}
\end{df}

Note how we decomposed our functor in two parts: one corresponding to the automaton structure, and the other one to the non-deterministic part. This is the same decomposition we already studied in section~\ref{coalgdet}. Our hope is to give a monadic structure to $\Pp \circ \Pp$, so that our definition of alternating automaton fits in the generic picture of determinization of a $\Tt$-automaton in section~\ref{coalgdet}.

\begin{eg}[Example of alternating automata]
The following automaton, with start state $q_0$, recognizes words with an even number of $a$ and $b$, or an odd number of $a$ and $b$.\\
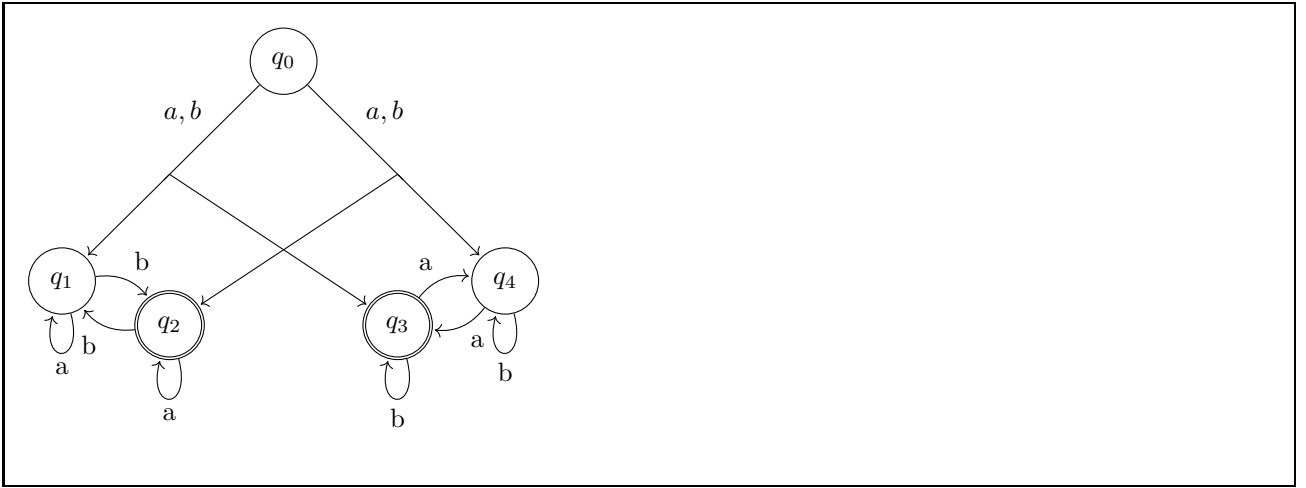
	\begin{tikzpicture}[shorten >=1pt,node distance=2cm,on grid,auto] 
		
		\node[state] (q_0)   {$q_0$}; 
		\node (f_1) at (-1.5,-1.5) {};
		\node (f_2) at (1.5,-1.5) {};
		\node[state] (q_1) [below left=of f_1] {$q_1$};
		\node[state, accepting] (q_2) [below =of f_1] {$q_2$}; 
		\node[state, accepting](q_3) [below =of f_2] {$q_3$};
		\node[state] (q_4) [below right=of f_2] {$q_4$} ;
		
		\path[-] 
		(q_0) edge node [swap] {$a,b$} (-1.6,-1.6)
		(q_0) edge node {$a,b$} (1.6,-1.6) ;
		
		\path[->]
		(-1.5,-1.5) 	edge node {} (q_1) 
		edge node {} (q_3)
		(1.5,-1.5)		edge node {} (q_2)
		edge node {} (q_4)
		(q_1)	edge [bend left] node {b} (q_2)
		edge [loop below] node {a} ()
		(q_2)	edge [bend left] node {b} (q_1)
		edge [loop below] node {a} ()
		(q_3)	edge [bend left] node {a} (q_4)
		edge [loop below] node {b} ()
		(q_4)	edge [bend left] node {a} (q_3)
		edge [loop below] node {b} ()
		;
	\end{tikzpicture}
\end{eg}

\subsection{A Monadic Structure For Alternating Automata}
\label{monadst}

This section contains most of the original work of the report. For reasons of clarity, we did not include the proofs of the original theorems we state in the section. However we added them in appendix~\ref{proof}, so that the interested reader can still have a look at them.

\paragraph{}
As we want to give a monadic structure to $\Pp \Pp$, and we already know $\Pp$ has a monadic structure, we are looking for a way to compose monads. The following proposition is the usual way it is done. It involves a distributive law between monads, which is defined in a similar way to a distributive law as defined in section~\ref{semdet}, with two more properties for the compatibility of the law with the second monad.

\begin{prop}[Composition of monads]
	Given two monads $(\Tt_1, \eta^1, \mu^1)$ and $(\Tt_2, \eta^2, \mu^2)$ and a distributive law between monads $\lambda : \Tt_1 \Tt_2 \To \Tt_2 \Tt_1$, one can create a new composite monad, with the components:
	\begin{itemize}
		\item functorial part $\Tt = \Tt_2 \circ \Tt_1$,
		\item unit $\eta$ defined by the components
		\[ \eta_X : X \to \Tt(X) = \left( X \stackrel{\eta^1_X}{\longrightarrow} \Tt_1(X) \stackrel{\eta^2_{\Tt_1(X)}}{\longrightarrow} \Tt_2(\Tt_1(X)) \right) \]
		\item multiplication $\mu$ defined by the components 
		\[\mu_X : \Tt^2(X) \to \Tt(X) = \left( \Tt_2(\Tt_1(\Tt_2(\Tt_1(X)))) \stackrel{\Tt_2(\lambda_{\Tt_1(X)})}{\longrightarrow} \Tt_2^2(\Tt_1^2(X)) \stackrel{\mu^2_{\Tt_1^2(X)}}{\longrightarrow} \Tt_2(\Tt_1^2(X)) \stackrel{\Tt_2(\mu^1_{X})}{\longrightarrow} \Tt_2(\Tt_1(X)) \right) \]
	\end{itemize}
\end{prop}

Note that naturality of the components of both monads ensure that the order in which the units and multiplications are performed are irrelevant, for instance one could apply $\eta^2$ first and the $\eta^1$, the resulting $\eta$ would be the same.

\paragraph{}
One could then try to construct a suitable distributive law for powerset over powerset, that should somehow reflect the fact that the outside powerset is considered disjunctively and the inside one is considered conjunctively. The natural candidate is the following law:
\[\begin{array}{rccl}
\lambda_X : & \Pp(\Pp(X)) & \to & \Pp(\Pp((X)) \\
& S & \mapsto & \{V \subseteq \cup S \mid \pta{U}{S} \operatorname{Card}(V \cap U) = 1 \}
\end{array}\]
that corresponds to the conversion of a disjunctive normal form into a conjunctive normal form.

However this is sadly not even a natural transformation. Patching it to
\[\begin{array}{rccl}
\lambda_X : & \Pp(\Pp(X)) & \to & \Pp(\Pp((X)) \\
& S & \mapsto & \{V \subseteq \cup S \mid \pta{U}{S} \operatorname{Card}(V \cap U) \geq 1 \}
\end{array}\]
(instead of taking exactly one element in every set, we take at least one) yields a natural transformation, however this natural transformation is not a distributive law (of monads over monads), and actually there are examples in \cite{NotMon} showing the natural transformation does not lead to a monad structure on $\Pp \Pp$.

A way to explain it is to observe that in the functor $\Pp \Pp$ we wish to turn into a monad, the outside $\Pp$ is in some way too relaxed. Indeed, if we have two forks $f \subseteq f' \subseteq X$, then the fork $f'$ accepts a word only if the fork $f$ accepts that word as well, so $\{f, f'\}$ accepts the exact same words as $\{f\}$. This is not very annoying for our concrete definition of acceptance, but as we just saw, when we move to category theory, the outside $\Pp$ causes troubles.

There are at least two different ways to patch this and get a unique set of fork to represent an acceptance condition:
\begin{itemize}
	\item require that no two forks are comparable; this replaces $\Pp(\Pp(X))$ by $\Aa(\Pp(X))$, the set of antichains (set of pairwise incomparable elements) of $\Pp(X)$,
	\item close the set of forks with respect to inclusion; this replaces $\Pp(\Pp(X))$ by $\tilde{\Up}(\Pp(X))$, the set of upward closed sets of $\Pp(X)$.
\end{itemize}

These two solutions are actually really close, as for any ordered set, there is a bijection between antichains over the set and upsets over the set (taking the minimal elements of an upset, and the upwards closure of an antichain). This is why we only explore one of the two idea — the one using upsets. It will enable us to give a correct distributive law, and following, also a correct monadic structure. This is why we alter the coalgebraic definition of alternating automata to the following:

\begin{df}[Alternating automata]
	An alternating automaton with respect to an alphabet $A$ is a coalgebra for the functor $\aut \circ \UUp \circ \Pp$, where $\Pp$ is the powerset monad (with codomain the category $\Poset$), and $\UUp$ is a functor from the category $\Poset$ to the category $\Set$ that takes a poset to the set of its upwards closed sets.\\
	We will denote the functor $\UUp \circ \Pp$ by $\Alt$.
\end{df}

Now, if we want to make this definition work with the results from section~\ref{coalgdet}, we need to give $\Alt$ a monadic structure. This is where category theory proves useful, because it will enable us to combine simple constructions in a reasonably easy way.

\begin{df}[Upwards closure, downwards closure]
	Given a poset $X$ and a subset $P$ of $X$, the upwards closure of $P$, denoted as $\upar P$ is the set defined by
	\[\upar P = \{x \in X \mid \exa{y}{P} y \leq x \} \]
	
	The downwards closure of $P$, denoted as $\downar P$, is defined similarly as
	\[\downar P = \{x \in X \mid \exa{y}{P} y \geq x \} \]
\end{df}

Note that the upwards (resp.\ downwards) closure of a set is always upwards (resp.\ downwards) closed.

Using this, we can define two monads on $\Poset$.
\begin{df}[Upset monad]
	We define the monad $(\Up, \eta^{\Up}, \mu^{\Up})$ on the category of posets as follows:
	\begin{itemize}
		\item on objects, $\Up((X,\preceq))$ is the set of upwards closed sets of $X$, ordered by \emph{reversed} inclusion order, that is $\Up(X) = (\{P \subseteq X \mid \pta{x,y}{X} x \preceq y \wedge x \in P \imp y \in P \}, \supseteq)$,
		\item on arrows, $\Up(f)(P) = \upar f(P) = \{y \in X \mid \exa{x}{P} f(x) \leq y \}$,
		\item the unit is $\eta^{\Up}(x) = \upar \{x\}$,
		\item the multiplication is $\mu^{\Up}(S) = \cup S$, as for the powerset monad.
	\end{itemize}
\end{df}

\begin{df}[Downset monad]
	We define the monad $(\Dn, \eta^{\Dn}, \mu^{\Dn})$ on the category of posets as follows:
	\begin{itemize}
		\item on objects, $\Dn(X)$ is the set of downwards closed sets of $X$, ordered by inclusion order, that is $\Dn(X) = (\{P \subseteq X \mid \pta{x,y}{X} x \geq y \wedge x \in P \imp y \in P \}, \subseteq)$,
		\item on arrows, $\Dn(f)(P) = \downar f(P)$,
		\item the unit is $\eta^{\Dn}(x) = \downar \{x\}$,
		\item the multiplication is $\mu^{\Dn}(S) = \cup S$.
	\end{itemize}	
\end{df}

The reason of the reversion of the order for the inclusion can be understood with the two following diagrams:\\
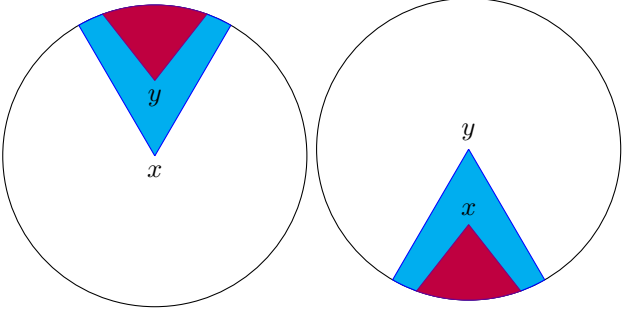
	\begin{tikzpicture}
	\draw (0,0) circle [radius = 2cm] ;
	\filldraw[fill=cyan, draw=blue] (0,0) node[below] {$x$} -- (60:2cm) arc (60:120:2cm) -- (0,0) ;
	\filldraw[fill=purple, draw=violet] (90:1cm) node[below] {$y$} -- (70:2cm) arc (70:110:2cm) -- (90:1cm) ;
	\end{tikzpicture}
	\begin{tikzpicture}
	\draw (0,0) circle [radius = 2cm] ;
	\filldraw[fill=cyan, draw=blue] (0,0) node[above] {$y$} -- (240:2cm) arc (240:300:2cm) -- (0,0) ;
	\filldraw[fill=purple, draw=violet] (270:1cm) node[above] {$x$} -- (250:2cm) arc (250:290:2cm) -- (270:1cm) ;
	\end{tikzpicture}\\
	On the leftmost, $x \leq y$ but $\uparrow x \supseteq \uparrow y$, whereas on the rightmost $x \leq y$ and $\uparrow x \subseteq \uparrow y$.

\paragraph{}
Note that the $\Up$ functor is of type $\Up : \Poset \to \Poset$, so the functor $\tilde{\Up}$ that we use in the definition of alternating automata is really just $\Uu \circ \Up$.

We also have a relation between $\Dn$, $\Do$ and $\Pp$:

\begin{prop}[Downwards closed sets of a discrete order]
	We have the equality
	\[\Dn \circ \Do = \Pp \]
\end{prop}

\begin{dem}
	Take a set $X$. By definition of $\Dn$, it holds that $\Dn \circ \Do(X)$ is a subset of $\Pp(X)$, ordered by inclusion. So we only need to show that any subset of $X$ is a downwards closed set of $\Do(X)$. Now, take $P$ a subset of $X$, $x$ and $y$ elements of $X$ and suppose $x \in P$. Because of the definition of the discrete order, if $x \geq y$ then $x = y$, and so $y \in P$. So $P$ is indeed downwards closed on $\Do(X)$.
\end{dem}

Thus, we can now rewrite the functor $\Alt$ as $\Uu \circ \Up \circ \Dn \circ \Do$. This does not seem much of an improvement, but we can now use the adjunction $\Do \dashv \Uu$ mentioned in section \ref{advconstr} and the monadic composition of $\Up$ and $\Dn$ to turn $\Alt$ into a monad. But first, we need a last ingredient. This is the distributive law we have been advertising throughout the paper. It has not been created for this paper, although tracing it back is hard to do. In our case, we found its description in \cite[p.~220-221]{BasDL}.

\begin{thm}[Distributive law]
	The following $\lambda : \Dn \Up \To \Up \Dn$ is a distributive law between monads:
	\[\begin{array}{rccl}
	\lambda_X :& \Dn(\Up(X)) & \to & \Up(\Dn(X)) \\
	& S & \mapsto & \{T \in \Dn(X) \mid \pta{s}{S} s \cap T \neq \emptyset \}
	\end{array} \]
\end{thm}

This transformation is similar to the way one transforms a disjunctive normal form into a conjunctive normal form in logic: to form a disjunction of conjunction from a conjunction of disjunction, one makes a big disjunction of all the different ways to pick one literal in each disjunction of the disjunctive normal form. Here, because of the upwards and downwards closure, at least one literal is taken rather than exactly one, but the idea is the same. Also, the functors $\Up$ and $\Dn$ encode this logical view on the powerset directly in the type, by making a clear difference between the monad interpreted disjunctively and the one interpreted conjunctively.

\paragraph{}
Using the first proposition of this section about the composition of monads, this distributive law yields a monad structure.

\begin{cor}[Monad structure for $\Up \Dn$]
	The functor $\Up \Dn$ can be given a monad structure $(\Up \Dn, \eta^1, \mu^1)$.
\end{cor}

This distributive law is the most important element of this report: giving a categorical semantics to alternating automata amounts to give a monad structure to some functor (either $\Pp \Pp$ or a modified version of it), which in turn amounts to finding a correct distributive law. Quite a lot of errors have been made trying to define this distributive law: a list appears in \cite{ErrMon}, itself being a correction of an error in \cite{ForgLog15}, tracing back the error to reference books such as \cite{MonSet}. A lot of patches have been found, but ours is the first completely satisfactory one: it preserves the properties of powerset (idempotency, commutativity, associativity), which was not the case in certain attempts (for instance, the lists/languages used in \cite{BialgDeterm}), while keeping the full power of alternating automata (contrary to the simpler version covered in \cite{TraceSem}), and it is an actual distributive law between monads (unlike the one proposed in \cite{ForgLog16}, a corrected version of \cite{ForgLog15}).

The main idea that had not been used before is to define the distributive law on $\Poset$ instead of $\Set$, and to use the adjunction $\Do \dashv \Uu$ to turn the obtained monad into a monad on $\Set$.

Using another category and adjunctions to construct a correct monad structure for alternating automata is not a new idea: an idea of how to do this (using semi-lattices and distributive lattices) is given in \cite{NotMon}. We tried this before devising our current solution, but the attempt was not successful.

\paragraph{}
The exact definition of the complete monad is then as follows:
\begin{thm}[Monad structure for $\Alt$]
	Let $\eta^2$ (resp.\ $\epsilon^2$) be the unit (resp.\ counit) of the adjunction $\Do \dashv \Uu$. Then the functor $\Alt$ (equal to $\UUp \Pp$ or $\Uu \Up \Dn \Do$) is a monad, which unit $\eta$ has components\\
	\begin{tikzcd}
			\eta_X : \id_X \arrow[r, "\eta^2_X"]& \Uu(\Do(X)) \arrow[r,"\Uu(\eta^1_{\Do(X)})"]& \Alt(X)
	\end{tikzcd}\\
	and multiplication $\mu$ has components\\
	\begin{tikzcd}[column sep = huge]
		\mu_X : \Alt(\Alt(X)) \arrow[r,"\Uu \Up \Dn \epsilon^2_{\Up \Dn \Do(X)}"] & \Uu \Up \Dn \Up \Dn \Do(X) \arrow[r, "\Uu(\mu^1_{\Do(X)})"] & \Alt(X) 
	\end{tikzcd}
\end{thm}

This construction is not ad-hoc, but it comes from the link between monads and adjunctions. See appendix~\ref{monadstructalt} for details.

One can easily compute the unit of this monad, and get
\[\eta_X(x) = \upar \{x\} = \{ T \in \Pp X \mid x \in T \} \]
For the multiplication, the computation gives the following:
\[\mu_X(S) = \{T \in \Pp(X) \mid \exa{s}{S} \pta{t}{s} \exa{u}{t} \pta{v}{u} v \in T \} \]
Because we constructed this multiplication stepwise, we can also give an intuition of how it works:
\begin{enumerate}
	\item use $\epsilon^2$ to get rid of the $\Do \Uu$ in the middle without really modifying the object (recall from section~\ref{advconstr} that $\epsilon^2$ is merely just the identity),
	\item use the distributive law to exchange the position of $\Up$ and $\Dn$, similarly as the transformation of a disjunctive normal form into a conjunctive normal form,
	\item flatten two levels of $\Up$ into one and two levels of $\Dn$ into one using the union.
\end{enumerate}

\subsection{Induced Semantics}
\label{indsem}

Now that we have a monad, we need an algebra for this monad to be able to define a semantics as in section~\ref{coalgdet}.

\begin{prop}[Algebra for $\Alt$]
	The pair $(\two, \beta)$ where
	\[\begin{array}{rccl}
	\beta : & \Alt(\two) & \to & \two \\
	& S & \mapsto & \left\{\begin{array}{l}
	\one \text{ if } \{\one\} \in S \\
	\zero \text{ otherwise}
	\end{array} \right.
	\end{array}\]
	is an algebra for $\Alt$.
\end{prop}

On a side note, one has $\Alt(\emptyset) = \UUp(\Pp(\emptyset)) = \UUp(\{\emptyset\}) = \{\emptyset, \{\emptyset \} \} = \two$, and the algebra $(\two,\beta)$ we just gave is actually $(\Alt(\emptyset), \mu_{\emptyset})$, which is usually called the free algebra on $\emptyset$.

\paragraph{}
Given this $\beta$ and a alternating automaton $(X,\langle o, \delta \rangle)$, there is a unique map $beh_1 : X \to \two^{A^*}$ that makes the following diagram commute:\\
\begin{tikzcd}[column sep = huge]
	X \arrow[r, "\beh_1"] \arrow[dd, "\langle o {,} \delta \rangle"'] & \two^{A^*} \arrow[d, "\langle O {,} D \rangle"] \\
	& \Dd(\two^{A^*})  \\
	\two \times (\Alt(X))^A = \Dd(\Alt(X)) \arrow[r, "\Dd(\Alt(\beh_1))"] & \Dd(\Alt(\two^{A^*})) \arrow[u,"\Dd(\hat{\beta})"']
\end{tikzcd}\\

If we spell out $\hat{\beta}$, we get $\hat{\beta}(S)(w) = \one \equ \exa{s}{S} \pta{L}{S} L(w) = \one$, and we can translate the above diagram into two conditions, corresponding to the first and second components of the products:
\begin{itemize}
	\item for $q \in X$, $\beh_1(q)(\emptyword) = o(q)$, so $\beh_1(q)$ accepts the empty word if and only if $o(q)$ is $\one$,
	\item for $q \in X$, $a \in A$ and $w \in A^*$, $\beh_1(q)(a \cdot w) = \one \equ \exa{F}{\delta(q)(a)} \pta{q'}{F} \beh_1(q')(w) = \one$.
\end{itemize}
This corresponds exactly to the concrete definition of acceptance that we gave for an alternating automaton in section~\ref{altaut}! So our whole construction is sound, as the categorical approach to alternating automata is equivalent to the usual one.

\section{Conclusion}

Using the construction described in this report, we are finally able to fully fit alternating automata in the large picture of determinization, and show that the problem there was not a failure of the theory, but rather a lack of a proper monad. A problem that we solved by finding a happy detour through order structures.

A nice thing about this construction is that it really shows the power of category theory: without it, formulating just the right monad would have been really hard. But using categorical tools like distributive law and adjunctions, breaking the problem in small, handleable pieces, and then putting these pieces together, makes the problem reasonable.

\phantomsection
\addcontentsline{toc}{section}{References}
\printbibliography

\newpage
\appendix
\appendixpage
\addappheadtotoc

\section{Proofs}
\label{proof}

\tikzcdset{row sep/normal=4em, column sep/normal =4em}

Because of the reversed inclusion order used in the monad $\Up(P)$, given a set of states $X$ ordered with some order $\leq$ and a subset $P$ of $X$, we need to make a difference between the upwards (resp.\ downwards) closure of $P$ with respect to $\leq$, and the upwards (resp.\ downwards) closure of $P$ with respect to the usual inclusion order. We will use the arrows $\upar$ (resp.\ $\downar$) for the first one, and $\uphar$ (resp.\ $\downhar$) for the second.

Also, we will write $\left\{x \in S \mid P(x) \right\}$ for the set of elements of $S$ having the property $P$ and $\left\{f(s), s \in S \right\}$ for the set $\left\{x \in T \mid \exa{s}{S} t = f(s)  \right\}$ if $T$ is the codomain of $f$.

Finally, to avoid confusion due to the many levels of intricate sets, we will write $f_*$ for the direct image, that is $f_*(S) = \left\{ f(s), s \in S\right\}$.

\subsection{Preliminary Order Lemmas}

\begin{lem}[Downwards closure and union]
	\label{closun}
	Given a set of sets $S$, we have $\cup \downhar S = \cup S$.
\end{lem}

\begin{lem}[Order closure and direct image]
	\label{closim}
	Given two posets $(X,\leq)$ and $(Y, \preceq)$, a monotone map $f : X \to Y$ and a set $P \subseteq X$, we have $\upar f_*(\upar P) = \upar f_*(P)$, and similarly $\downar f_*(\downar P) = \downar f_*(P)$.
\end{lem}

\begin{lem}[Upwards closure and intersection]
	\label{closint}
	If $S$ is a set of sets and $t$ is a set, then $\pta{s}{\uphar S} t \cap s \neq \emptyset$ is equivalent to $\pta{s}{S} t \cap s \neq \emptyset$.
\end{lem}

\subsection[Monad Structure of Up and Dn]{Monad Structure of $\Up$ and $\Dn$}

\begin{prop}[Upset monad]
	The triple $(\Up, \eta^{\Up}, \mu^{\Up})$, defined as follows:
	\begin{itemize}
		\item on objects, $\Up((X\leq))$ is the set of upwards closed sets of $X$, ordered by \emph{reversed} inclusion order, that is $\Up((X,\leq)) = (\left\{P \subseteq X \mid \pta{x,y}{X} x \leq y \wedge x \in P \imp y \in P \right\}, \supseteq)$,
		\item on arrows, $\Up(f)(P) = \upar f_*(P)$,
		\item the unit is $\eta^{\Up}(x) = \upar \left\{x\right\}$,
		\item the multiplication is $\mu^{\Up}(S) = \cup S$,
	\end{itemize}
	is a monad.
\end{prop}

\begin{dem}
	In this proof we write $\eta$ for $\eta^{\Up}$ and $\mu$ for $\mu^{\Up}$.
	
	First, we need to prove that $\Up$ is a functor. Since direct image and upwards closure preserve inclusion, if $f : X \to Y$ is an arrow in $\Poset$, then $\Up(f)$ is monotone, and so it is an arrow $\Up(X) \to \Up(Y)$. Checking that $\Up$ preserves identity is easy. Finally, if $f$ and $g$ are two arrows with the correct types, $\upar g_*(\upar f_*(S)) = \upar g_*(f_*(S)) = \upar (g \circ f)_*(S)$, using lemma~\ref{closim}.
	
	Next, the unit. The upwards closure of the image by $\eta$ is obvious. Next, suppose $(X,\leq)$ is a poset, and $x, y \in X$ are such that $x \leq y$. Then $y \in \upar \left\{x\right\}$, so $\upar \left\{y \right\} \subseteq \upar \left\{x \right\}$, and so $\eta_{(X,\leq)}(x) \supseteq \eta_{(X,\leq)}(y)$. But because $\Up(X)$ is ordered with respect to the \emph{reversed} inclusion order, $\eta_{(X,\leq)}$ is monotone, and it is an arrow $\eta_{P} : P \to \Up(P)$. To prove the naturality, we need to show $\upar f_*(\upar \left\{x\right\}) = \upar \left\{f(x)\right\}$. But using lemma~\ref{closim}, we have $\upar f_*(\upar \left\{x\right\}) = \upar f_*(\left\{x\right\})$ and since $f_*(\left\{x\right\}) = \left\{f(x)\right\}$, we have naturality of $\eta$. 
	
	Now, the multiplication. Any union of upwards closed sets is upwards closed, and if $S \subseteq S'$ then $\cup S \subseteq (\cup S) \cup (\cup (S' \backslash S)) = \cup S'$ so $\mu_P$ is an arrow $\mu_P : \Up(\Up(P)) \to \Up(P)$. For naturality, we need to prove $\cup (\upar \left\{\upar f_*(s), s \in S \right\}) = \upar f_*(\cup S)$. Now because of the reversed order on $\Up(P)$, we have $\cup (\upar \left\{\upar f_*(s), s \in S \right\}) = \cup (\downhar \left\{\upar f_*(s), s \in S \right\})$, and using lemma~\ref{closun}, we deduce
	\begin{align*}
		\cup (\upar \left\{\upar f_*(s), s \in S\right\}) &= \cup \left\{\upar f_*(s), s \in S  \right\} \\
		&= \cup \left\{\upar \left\{f(x), x \in s\right\}, s \in S  \right\} \\
		&= \left\{\upar f(x), x \in \cup S \right\} \text{ because $x \in s \in S \equ x \in \cup S$}
	\end{align*}
	and so $\mu$ is natural.
	
	Next, the multiplication and unit are compatible, because $\mu_X \circ \eta_{\Up(X)}(S) = \cup \upar \left\{ S\right\} = \cup \downhar \left\{S \right\} = S$ and $\mu_X \circ \Up(\eta_X)(S) = \cup_{s \in S}(\upar \left\{s\right\})$, so $x \in \mu \circ \Up(\eta_X)(S)$ if and only if $\exa{s}{S} x \geq s$, that is if and only if $x \in \upar S$. But since $S$ is an upset, $\upar S = S$. So we indeed have\\
	
	\begin{tikzcd}
		\Up X \arrow[r, "\eta_{\Up X}"] \arrow[d, "\Up(\eta_X)"'] \arrow[dr, equal] & \Up \Up X \arrow[d, "\mu_X"] \\
		\Up \Up X \arrow[r,"\mu_X"'] & \Up(X)
	\end{tikzcd}\\
	
	Finally the diagram\\
	\begin{tikzcd}
		\Up \Up \Up X \rar["\Up(\mu_X)"] \dar["\mu_{\Up X}"] & \Up \Up X \dar["\mu_X"] \\
		\Up \Up X \rar["\mu_X"] & \Up X
	\end{tikzcd}\\
	is easy to deduce, as
	\begin{align*}
		\cup \upar \left\{\cup s, s \in S \right\} &= \cup \downhar \left\{ \cup s, s \in S\right\} \\
		&= \cup \left\{\cup s, s \in S \right\} \\
		&= \left\{x \in X \mid \exa{s}{S} \exa{t}{s} x \in t \right\} \\
		&= \cup \cup S
	\end{align*}
	
\end{dem}

\begin{prop}[Downset monad]
	The triple $(\Dn, \eta^{\Dn}, \mu^{\Dn})$, defined as follows:
	\begin{itemize}
		\item on objects, $\Dn(X)$ is the set of downwards closed sets of $X$, ordered by inclusion order, that is $\Dn(X) = (\left\{P \subseteq X \mid \pta{x,y}{X} x \geq y \wedge x \in P \imp y \in P \right\}, \subseteq)$,
		\item on arrows, $\Dn(f)(P) = \downar f_*(P)$,
		\item the unit is $\eta^{\Dn}(x) = \downar \left\{x\right\}$,
		\item the multiplication is $\mu^{\Dn}(S) = \cup S$,
	\end{itemize}
	is a monad.
\end{prop}

\begin{dem}
	The proof is extremely similar to the one for $\Up$, so we will not repeat it. The only thing to note is that every time we transformed a $\upar$ into a $\downhar$ in the previous proof, in the case of $\Dn$ we would get a $\downar$ on a set ordered by regular inclusion, so it would also translate into $\downhar$.
\end{dem}

\subsection{Naturality Of The Distributive Law}

\begin{thm}[Distributive law]
	The following $\lambda : \Dn \Up \To \Up \Dn$ is a distributive law between monads:
	\[\begin{array}{rccl}
	\lambda_X :& \Dn(\Up(X)) & \to & \Up(\Dn(X)) \\
	& S & \mapsto & \left\{T \in \Dn(X) \mid \pta{s}{S} s \cap T \neq \emptyset \right\}
	\end{array} \]
\end{thm}

\begin{dem}
	First, $\lambda$ is well defined because $\left\{T \in \Dn(X) \mid \pta{s}{S} s \cap T \neq \emptyset \right\}$ is an upset of downsets.\\
	
	Next, we prove naturality. Take $X, Y$ posets and $f : X \to Y$ and arrow in $\Poset$. We need to prove the following diagram:\\
	\begin{tikzcd}
		\Dn \Up X \rar["\lambda_X"] \dar["\Dn \Up f"] & \Up \Dn X \dar["\Up \Dn f"] \\
		\Dn \Up Y \rar["\lambda_Y"] & \Up \Dn Y
	\end{tikzcd}\\
	Take $S \in \Dn \Up X$ and $d \in \Dn Y$. First, we rewrite the lower triangle:
	\begin{align*}
		d \in \lambda_Y \circ \Dn \Up(f)(S)
		&\equ \pta{u}{\Dn \Up f (S)} u \cap d = \emptyset\\
		&\equ \pta{u}{\downar \left\{\upar f_*(s), s \in S \right\}} u \cap d \neq \emptyset\\
		&\equ \pta{u}{\uphar \left\{\upar f_*(s), s \in S\right\}} u \cap d \neq \emptyset \\
		&\equ \pta{u}{\left\{\upar f_*(s), s \in S\right\}} u \cap d \neq \emptyset \text{ by lemma~\ref{closint}} \\
		&\equ \pta{s}{S} d \cap \upar f_*(s) \neq \emptyset \\
		&\equ \pta{s}{S} d \cap f_*(s) \neq \emptyset \text{ because $d$ is a downwards closed set} \\
		&\equ \pta{s}{S} \exa{x}{s} f(x) \in d
	\end{align*}
	Now the upper triangle, we obtain:
	\begin{align*}
		d \in \Up \Dn(f) \circ \lambda_X(S)
		& \equ d \in \upar \left\{\downar f_*(t), t \in \lambda_X(S) \right\}\\
		& \equ \exa{d'}{\Dn Y} d \geq d' \wedge \exa{t}{\lambda_X(S)} d' = \downar f_*(t) \\
		& \equ \exa{t}{\Dn X} (\pta{s}{S} t \cap s \neq \emptyset) \wedge (d \geq  \downar f_*(t))\\
		& \equ \exa{t}{\Dn X} (\pta{s}{S} t \cap s \neq \emptyset) \wedge (d \supseteq \downar f_*(t)) \\
		& \equ \exa{t}{\Dn X} (\pta{s}{S} t \cap s \neq \emptyset) \wedge (f_*(t) \subseteq d) \text{ because $d$ is a downset} \\
		& \equ \exa{t}{\Dn X} (\pta{s}{S} t \cap s \neq \emptyset) \wedge (\pta{x}{t} f(x) \in d)
	\end{align*}
	Now suppose $\exa{t}{\Dn X} (\pta{s}{S} t \cap s \neq \emptyset) \wedge (\pta{x}{t} f(x) \in d)$. Then given $s \in S$, there is some $x_s \in t \cap s$, but then $f(x_s) \in d$, and so $\pta{s}{S} \exa{x}{s} f(x) \in d$. Conversely, suppose $\pta{s}{S} \exa{x}{s} f(x) \in d$. For each $s \in S$ fix some $x_s \in s$ such that $f(x_s) \in d$, and define $t = \downar \left\{x_s, s \in S \right\}$. Then by construction $t \in \Dn X$, and moreover $\pta{s}{S} t \cap s \supseteq \left\{ x_s\right\} \supsetneq \emptyset$. Also, if $x \in t$, then take some $s \in S$ such that $x \leq x_s$. Then $f(x) \leq f(x_s) \in d$, and because $d$ is a downwards closed set, $f(x) \in d$. Thus, $\exa{t}{\Dn X} (\pta{s}{S} t \cap s \neq \emptyset) \wedge (\pta{x}{t} f(x) \in d)$.\\	
	In the end $d \in \lambda_Y \circ \Dn \Up(f)(S) \equ d \in \Up \Dn(f) \circ \lambda_X(S)$ and so the diagram commutes.
		
\paragraph{}	
	Next, the first triangle diagram, namely\\
	\begin{tikzcd}
		\Up X \rar["\eta^{\Dn}_{\Up X}"] \drar["\Up(\eta^{\Dn}_X)"'] & \Dn \Up X \dar["\lambda_X"] \\
		& \Up \Dn X
	\end{tikzcd}\\
	Take $S \in \Up X$, we have one one hand
	\begin{align*}
		\Up(\eta^{\Dn}_{X})(S) &= \upar \left\{ \downar s, s \in S\right\} \\
		&= \left\{T \in \Dn X \mid \exa{s}{S} \downar s \subseteq T \right\} \\
		&= \left\{T \in \Dn X \mid \exa{s}{S} s \in T \right\} \text{ because $T$ is a downwards closed set} \\
		&= \left\{T \in \Dn X \mid T \cap S \neq \emptyset \right\}
	\end{align*}
	On the other hand,
	\begin{align*}
		\lambda_X \circ \eta^{\Dn}_{\Up X}(S)
		&= \left\{T \in \Dn X \mid \pta{s}{\eta^{\Dn}_{\Up X}(S)} s \cap T \neq \emptyset \right\} \\
		&= \left\{T \in \Dn X \mid \pta{s}{\downar \left\{S \right\}} s \cap T \neq \emptyset \right\} \\
		&= \left\{T \in \Dn X \mid \pta{s}{\uphar \left\{S \right\}} s \cap T \neq \emptyset \right\} \\
		&= \left\{T \in \Dn X \mid \pta{s}{\left\{S\right\}} s \cap T \neq \emptyset \right\} \text{ by lemma~\ref{closint}} \\
		&= \left\{T \in \Dn X \mid S \cap T \neq \emptyset \right\} 
	\end{align*}
	and so the diagram commutes.

\paragraph{}
	The second triangle diagram is:\\
	\begin{tikzcd}
		\Dn X \drar["\eta^{\Up}_{\Dn X}"'] \rar["\Dn(\eta^{\Up}_X)"] & \Dn \Up X \dar["\lambda_X"] \\
		& \Up \Dn X
	\end{tikzcd}\\
	Take $S \in \Dn X$, we have on one hand
	\begin{align*}
		\eta^{\Up}_{\Dn X}(S) &= \upar \left\{ S\right\}
	\end{align*}
	On the other hand, we have
	\begin{align*}
		\lambda_X \circ \Dn(\eta^{\Up}_{X})(S)
		&= \left\{T \in \Dn(X) \mid \pta{t}{\downar \left\{\upar \left\{s \right\}, s \in S \right\}} T \cap t \neq \emptyset \right\} \\
		&= \left\{T \in \Dn(X) \mid \pta{t}{\uphar \left\{\upar \left\{s \right\}, s \in S \right\}} T \cap t \neq \emptyset \right\} \\
		&= \left\{T \in \Dn(X) \mid \pta{t}{\left\{\upar \left\{s \right\}, s \in S \right\}} T \cap t \neq \emptyset \right\} \text{ by lemma~\ref{closint}} \\
		&= \left\{T \in \Dn(X) \mid \pta{s}{S} T \cap \upar \left\{s\right\} \neq \emptyset \right\} \\
		&= \left\{T \in \Dn(X) \mid \pta{s}{S} s \in T \right\} \text{ because $T$ is a downset} \\
		&= \left\{T \in \Dn(X) \mid S \subseteq T \right\} \\
		&= \upar \{S\}
	\end{align*}
	And so this triangle diagram commutes as well.

\paragraph{}
	Now, the first rectangle diagram, that is\\
	\begin{tikzcd}
		\Dn \Dn \Up X \rar["\Dn(\lambda_X)"] \dar["\mu^{\Dn}_{\Up X}"] & \Dn \Up \Dn X \rar["\lambda_{\Dn X}"] & \Up \Dn \Dn X \dar["\Up(\mu^{\Dn}_X)"] \\
		\Dn \Up X \arrow[rr, "\lambda_X"] && \Up \Dn X
	\end{tikzcd}\\
	Take $S \in \Dn \Dn \Up X$, on one hand we have
	\begin{align*}
		\lambda_X \circ \mu^{\Dn}_{\Up X}(S)
		&= \left\{T \in \Dn X \mid \pta{s}{\cup S}, s \cap T \neq \emptyset \right\} \\
		&= \left\{T \in \Dn X \mid \pta{s}{S} \pta{t}{s} t \cap T \neq \emptyset \right\}
	\end{align*}
	On the other hand, we have
	\begin{align*}
		\Up(\mu^{\Dn}_X) \circ \lambda_{\Dn X} \circ \Dn(\lambda_X)(S)
		&= \upar \left\{\cup T, T \in \lambda_{\Dn X}(\Dn(\lambda_X)(S)) \right\} \\
		&= \upar \left\{\cup T, T \in \left\{T \in \Dn \Dn X \mid \pta{s}{\Dn(\lambda_X)(S)} s \cap T \neq \emptyset \right\}  \right\} \\
		&= \upar \left\{\cup T, T \in \left\{T \in \Dn \Dn X \mid \pta{s}{\downar(\lambda_X)_*(S)} s \cap T \neq \emptyset \right\}  \right\} \\
		&= \upar \left\{\cup T, T \in \left\{T \in \Dn \Dn X \mid \pta{s}{\uphar(\lambda_X)_*(S)} s \cap T \neq \emptyset \right\}  \right\} \\
		&= \upar \left\{\cup T, T \in \left\{T \in \Dn \Dn X \mid \pta{s}{(\lambda_X)_*(S)} s \cap T \neq \emptyset \right\}  \right\} \text{ by lemma~\ref{closint}} \\
		&= \upar \left\{\cup T, T \in \left\{ T \in \Dn \Dn X \mid \pta{s}{S} \lambda_X(s) \cap T \neq \emptyset \right\} \right\} \\
		&= \upar \left\{\cup T, T \in \left\{T \in \Dn \Dn X \mid \pta{s}{S} T \cap \left\{u \in \Dn X \mid \pta{t}{s} t \cap u \neq \emptyset \right\} \neq \emptyset \right\}  \right\} \\
		&= \upar \left\{\cup T, T \in \left\{T \in \Dn \Dn X \mid \pta{s}{S} \exa{u}{T} \pta{t}{s} t \cap u \neq \emptyset \right\}  \right\}
	\end{align*}
	Take $T \in \Dn X$ and suppose $T \in \lambda_X \circ \mu^{\Dn}_{\Up X}(S)$, that is
	\[\pta{s}{S} \pta{t}{s} t \cap T \neq \emptyset\]
	Then we have
	\[\pta{s}{S} \exa{u}{(\downhar \left\{T\right\})} \pta{t}{s} t \cap u \neq \emptyset\]
	with $u$ taken each time to be $T$. But $\downhar \left\{ T\right\}$ is a downset, thus, $\downhar \left\{ T \right\} \in \left\{T \in \Dn \Dn X \mid \pta{s}{S} \exa{u}{T} \pta{t}{s} t \cap u \neq \emptyset \right\}$. Since $T = \cup \downhar \left\{ T\right\}$, we deduce that $T \in \Up(\mu^{\Dn}_X) \circ \lambda_{\Dn X} \circ \Dn(\lambda_X)(S)$.
	
	Conversely, take $T \in \Dn \Dn X$ such that $\pta{s}{S} \exa{u}{T} \pta{t}{s} t \cap u \neq \emptyset$, and take $s \in S$. By hypothesis, take $u_s \in T$ such that $\pta{t}{s} t \cap u_s \neq \emptyset$. Take some $t \in s$, we have $t \cap \cup T \supseteq t \cap u_s \supsetneq \emptyset$, and so $t \cap \cup T \neq \emptyset$. This proves $\cup T \in \lambda_X \circ \mu^{\Dn}_{\Up X}(S)$. Because the set $\lambda_X \circ \mu^{\Dn}_{\Up X}(S)$ is an upwards closed set, for any $T' \geq T$ we still have $T' \in \lambda_X \circ \mu^{\Dn}_{\Up X}(S)$. Thus, we have $\Up(\mu^{\Dn}_X) \circ \lambda_{\Dn X} \circ \Dn(\lambda_X)(S) \subseteq \lambda_X \circ \mu^{\Dn}_{\Up X}(S)$.
	
	The double inclusion proves the equality, and so the diagram commutes.

\paragraph{}
	Finally, the last diagram to prove is\\
	\begin{tikzcd}
		\Dn \Up \Up X \rar["\lambda_{\Up X}"] \dar["\Dn(\mu^{\Up}_X)"] & \Up \Dn \Up X \rar["\Up(\lambda_X)"] & \Up \Up \Dn X \dar["\mu^{\Up}_{\Dn X}"] \\
		\Dn \Up X \arrow[rr,"\lambda_X"] && \Up \Dn X
	\end{tikzcd}\\
	Take $S \in \Dn \Up \Up X$, the lower part gives
	\begin{align*}
		\lambda_X \circ \Dn(\mu^{\Up}_X)(S)
		&= \{ t \in \Dn X \mid \pta{u}{\Dn(\mu^{\Up}_X)(S)} u \cap t \neq \emptyset \} \\
		&= \{t \in \Dn X \mid \pta{u}{\downar \{ \cup s, s \in S\}} u \cap t \neq \emptyset \} \\
		&= \{t \in \Dn X \mid \pta{u}{\uphar \{\cup s, s \in S \}} u \cap t \neq \emptyset \} \\
		&= \{t \in \Dn X \mid \pta{u}{\{\cup s, s \in S \} } u \cap t \neq \emptyset \} \text{ by lemma~\ref{closint}} \\
		&= \{ t \in \Dn X \mid \pta{s}{S} t \cap \cup s \neq \emptyset \}
	\end{align*}
	The upper part gives
	\begin{align*}
		\mu^{\Up}_{\Dn X} \circ \Up(\lambda_X) \circ \lambda_{\Up X}(S)
		&=\cup \upar \{\lambda_X(T), T \in \lambda_{\Up X}(S) \} \\
		&= \cup \downhar \{\lambda_X(T), T \in \{T \in \Dn \Up X \mid \pta{s}{S} T \cap s \neq \emptyset \} \\
		&= \cup \{\lambda_X(T), T \in \{T \in \Dn \Up X \mid \pta{s}{S} T \cap s \neq \emptyset \} \text{ by lemma~\ref{closun}} \\
		&= \{t \in \Dn X \mid \exa{T}{\Dn \Up X} (\pta{s}{S} T \cap s \neq \emptyset)\wedge (\pta{x}{T} x \cap t \neq \emptyset) \}
	\end{align*}
	Now take $t \in \Dn X$, and suppose $\pta{s}{S} t \cap \cup s \neq \emptyset$. For each $s \in S$, fix $y_s \in s$ and $x_s \in y_s$ such that $x_s \in t$. Define $T = \uphar \{y_s, s \in S \} = \downar \{y_s, s \in S \}$, by definition we have $T \in \Dn \Up X$. Moreover, for any $s \in S$, we have $y_s \in s \cap T$ and so $\pta{s}{S} T \cap s \neq \emptyset$. Also, for a given $y \in T$, there is some $s \in S$ such that $y \supseteq y_s$. But then $t \cap y \supseteq t \cap y_s \supsetneq \emptyset$. Therefore, $t \in \mu^{\Up}_{\Dn X} \circ \Up(\lambda_X) \circ \lambda_{\Up X}(S)$.
	
	Conversely, take $t \in \Dn X$ and suppose $\exa{T}{\Dn \Dn X} (\pta{s}{S} T \cap s \neq \emptyset)\wedge (\pta{x}{T} x \cap t \neq \emptyset)$. Take such a $T$, and some $s \in S$. By hypothesis, there is a $y_s \in T \cap s$, and because $y_s \in T$, there is some $x_s \in y_s \cap t$. But then $x_s \in \cup s$, and so $t \cap \cup s \neq \emptyset$. Thus, $t \in \lambda_X \circ \Dn(\mu^{\Up}_X)(S)$.\\
	By double inclusion, the diagram commutes.
		
\end{dem}

\subsection[Monad Structure of Alt]{Monad Structure of $\Alt$}
\label{monadstructalt}

The construction we used to construct the monad $\Alt$ from a monad and an adjunction is not ad-hoc. Instead it is a consequence of the three following facts, which are classical results on monads and adjoints (see for instance \cite[chapter~4,~6]{McLane}):

\begin{thm}[Monad arising from an adjunction]
	\label{MonAdj}
	Given an adjunction $\Ff \vdash \Gg$ with unit $\eta$, the functor $\Gg \circ \Ff$ can be equipped with a monad structure, whose unit is the unit of the adjunction.
\end{thm}
The multiplication of the monad can also be described in term of the adjunction, but not in a simple way, so we leave this out.

\begin{prop}[Adjoints arising from a monad]
	\label{AdjMon}
	Given a monad $(\Tt,\eta,\mu)$ on a category $\CC$, the monad arises from the following adjunction:\\
	\begin{tikzcd}
		\CC \arrow[rr, bend left = 30, "\Ff"] & \bot & \Em(\Tt) \arrow[ll, bend left = 30, "\Uu"]
	\end{tikzcd}\\
	Where $\Em(T)$ is the category of algebras for $\Tt$, $\Ff(X) = (\Tt X, \mu_X)$, $\Ff(f) = T(f)$, $\Uu'(X, f) = X$ and $\Uu'(\alpha) = \alpha$.
\end{prop}

	The exact details of the two adjoint functors is not very relevant, the important part is that the monad arising from the adjunction is $\Tt$.

\begin{prop}[Composition of adjoints]
	\label{CompAdj}
	If $\Ff \dashv \Gg$ is an adjunction between $\CC$ and $\DD$, with unit $\eta$ and counit $\epsilon$ and $\Ff' \dashv \Gg'$ is an adjunction between $\DD$ and $\EE$ with unit $\eta'$ and counit $\epsilon'$, that is we are in the following situtation:\\
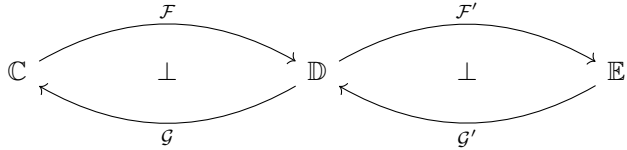
		\begin{tikzcd}
			\CC \arrow[rr, bend left = 30, "\Ff"] & \bot & \DD \arrow[ll, bend left = 30, "\Gg"]
			\arrow[rr, bend left = 30, "\Ff'"] & \bot & \EE \arrow[ll, bend left = 30, "\Gg'"] 
		\end{tikzcd}\\
		then $\Ff' \circ \Ff \dashv \Gg \circ \Gg'$ is an adjunction between $\CC$ and $\EE$, with unit
		\[\eta'' : \id_{\CC} \stackrel{\eta}{\To} \Gg \circ \Ff \stackrel{\Gg \eta' \Ff}{\To} \Gg \Gg' \Ff' \Ff \]
		and counit
		\[\mu'' : \Ff' \Ff \Gg \Gg' \stackrel{\Ff' \epsilon \Gg'}{\To} \Ff' \Gg' \stackrel{\epsilon'}{\To} \id_{\EE} \]
\end{prop}

Now the monadic structure for $\Alt$ can be constructed in three steps:
\begin{enumerate}
	\item from $\Up \Dn$ get an adjunction 	\begin{tikzcd}
		\Poset \arrow[rr, bend left = 30, "\Ff"] & \bot & \Em(\Up \Dn) \arrow[ll, bend left = 30, "\Uu'"]
	\end{tikzcd}\\using proposition~\ref{AdjMon}
	\item compose this adjunction with 	\begin{tikzcd}
	\Set \arrow[rr, bend left = 30, "\Do"] & \bot & \Poset \arrow[ll, bend left = 30, "\Uu"]
	\end{tikzcd}\\using proposition~\ref{CompAdj}
	\item from this composite adjunction, get a monad on $\Set$ using propoition~\ref{MonAdj}      
\end{enumerate}
And this monad on $\Set$ is exactly the monad on $\Alt$ described at the end of section~\ref{monadst}.

\section{Thanks}

I would like to thank Jurriaan Rot for being my guide in the world of coalgebras and pointing me again and again in the right direction, Luigi Santocanale whom I never met in person but who still gave me the right thing to look at for the distributive law, Joshua Moerman for his paper summing up all the troubles people have had on the same problem as me before, and Alexandre Goy, for keeping the mood up in the office.


\end{document}